\documentclass[english]{article}
\usepackage[T1]{fontenc}
\usepackage[latin9]{inputenc}
\usepackage[letterpaper]{geometry}
\usepackage{float}
\usepackage{units}
\usepackage{amsmath}
\usepackage{esint}

\makeatletter

\providecommand{\tabularnewline}{\\}

\makeatother

\usepackage{babel}

\begin{document}

\title{Benford's law: a theoretical explanation for base 2}

\author{H. M. Bharath, IIT Kanpur}
\maketitle
\begin{abstract}
In this paper, we present a possible theoretical explanation for benford's
law. We develop a recursive relation between the probabilities, using
simple intuitive ideas. We first use numerical solutions of this recursion
and verify that the solutions converge to the benford's law. Finally
we solve the recursion analytically to yeild the benford's law for
base 2. 
\end{abstract}

\section{Introduction}

The leading significant digit of a random integer is one of $1,2\cdots9$.
Intuitively, it is equally likely to be any of these nine figures.
However, empirical observations, and the benford's law indicate the
contrary. According to the law, the probability that a random integer,
expressed in base 10, starts with the digit $d$ is\cite{key-1} \begin{equation}
P_{d}=Log_{10}(1+\frac{1}{d})\end{equation}
$d=1,2\cdots9$. This law was first proposed by newcomb in 1881\cite{key-2}.
It means, a random integer is most likely to start with 1, with a
probability of 0.301, and least likely to start with 9, with a probability
of 0.046. Note that the random integer is $unscaled$; i.e., it can
be arbitrarily large. This is the suspected reason behind the nonuniform
probabilities. On the other hand, if the random number was $scaled$,
i.e, chosen from a bounded set, the corresponding probabilities are
obtained through a direct calculation. For instance, consider a scale
of $100$, ie, the number is chosen from the set $[0,100)$; the probabilities
are indeed uniform. However, if the scale were $200$, they would
be nonuniform, with $d=1$ acquiring a very large probability($>\frac{1}{2}$).
In this paper, we use these scaled probabilities to arrive at the
benford's values of unscaled probabilities. Before we proceed, we
shall state the well known generalizations of benford's law. 

The law is generalized to first two digits. The probability that a
random integer starts with digits $d_{1}d_{2}$ is given by \begin{equation}
P_{d_{1}d_{2}}=Log_{10}(1+\frac{1}{d_{2}+10d_{1}})=Log_{10}(1+\frac{1}{d_{1}d_{2}})\end{equation}
It is further generalised to arbitrary number of significant digits,
and expressed in an arbitrary base $b$ as

\begin{equation}
P_{d_{1}\cdots d_{k}}=Log_{b}(1+\frac{1}{d_{k}+bd_{k-1}+\cdots+b^{k-1}d_{1}})=Log_{b}(1+\frac{1}{d_{1}\cdots d_{k}})\end{equation}
where $d_{1}\cdots d_{k}$ is the number expressed in base $b$\cite{key-4}.
We shall consider the simple case of base $2$. In the next section,
we present the basic idea behind the proof, supported with examples
and numerical calculations. The analytical proof is provided in section
3. We end with a brief discussion, in section 4.

\section{Basic idea behind the proof and numerical estimates}

Expressed in base $2$, every number starts with $1$. Hence we consider
the first two significant digits, which are either $10$ or $11$.
Let $P_{10}$ and $P_{11}$ be the corresponding probabilities. According
to benford's law, $P_{10}=log_{2}(1+\frac{1}{2})=0.5849625$, $P_{11}=log_{2}(1+\frac{1}{3})=0.4150375$. 

These are the unscaled probabilities. Unlike them, the scaled probabilities
are easily evaluated. For instance, consider a scale of $1000$; i.e,
the random integer is chosen from the set $[0,1000)$. Since, in this
set, numbers starting from $10$ and $11$ are equally populated,
the corresponding probabilities are $\frac{1}{2}$ each. This is true
of any scale of the form $100\cdots0$. Accordingly let us denote
them by $P_{10}^{10}=P_{11}^{10}=\frac{1}{2}$. The superscript indicates
that the scale is of the form $10\cdots0$. Now consider a scale of
$1100$. It can be verified that the probabilities are now $\frac{2}{3}$
and $\frac{1}{3}$. Also, this is true of any scale of the form $110\cdots0$.
Let us denote them by $P_{10}^{11}=\frac{2}{3}$ and $P_{11}^{11}=\frac{1}{3}$. 

Thus, the unscaled probability $P_{10}$ is in between $P_{10}^{10}$
and $P_{10}^{11}$, and $P_{11}$ is inbetween $P_{11}^{10}$ and
$P_{11}^{11}$. \begin{equation}
P_{10}=P_{10}^{10}w+P_{10}^{11}(1-w)\end{equation}
\begin{equation}
P_{11}=P_{11}^{10}w+P_{11}^{11}(1-w)\end{equation}
where $w$ is the weight assosiated with the scale being of the form
$100\cdots0$. To a first order, it can be approximated to the probability
that a randomly chosen scale $starts$ $with$ $10$, which is $P_{10}$.
Thus,\begin{equation}
P_{10}=P_{10}^{10}P_{10}+P_{10}^{11}P_{11}\end{equation}
 \begin{equation}
P_{11}=P_{11}^{10}P_{10}+P_{11}^{11}P_{11}\end{equation}
this gives $P_{10}=\frac{4}{7}=0.57142$, and $P_{11}=\frac{3}{7}=0.42857$.
These are the first order approximations. The approximation lies in
the assumption $w=P_{10}$; all integers starting from $10$ are not
of the form $10\cdots0$. 

To sharpen the approximation, consider the first three significant
digits. Using a similar notation, we denote the unscaled probabilities
by $P_{1xy}$, where $x,y=0,1$. And the scaled probabilities by $P_{1xy}^{1\alpha\beta}$,
$x,y,\alpha,\ensuremath{\beta}=0,1$. $P_{1xy}^{1\alpha\beta}$ is
the probability that a random integer starts with $1xy$ when the
scale is of the form $1\alpha\beta0\cdots0$. The equations, to the
second approximation are\begin{equation}
P_{1xy}=\sum_{\alpha\beta}P_{1xy}^{1\alpha\beta}P_{1\alpha\beta}\end{equation}
This is a set of four equations in four variables. Once we solve for
$P_{1\alpha\beta}$, we can evaluate $P_{10}$ using $P_{10}=P_{100}+P_{101}.$

To do this, we are to first evaluate $P_{1xy}^{1\alpha\beta}$, the
population fraction of numbers starting from $1xy$ in the integer
set $S=[0,1\alpha\beta0\cdots0)$. This set can be broken in to three
chunks $S=S_{0}\cup S_{1}\cup S_{2}$ where $S_{0},S_{1}$ and $S{}_{2}$
are the integer sets,\begin{eqnarray*}
S_{0}=[0,1000\cdots0)\\
S_{1}=[1000\cdots0,1\alpha00\cdots0)\\
S_{2}=[1\alpha00\cdots0,1\alpha\beta0\cdots0)\end{eqnarray*}
Note that they are disjoint. $S_{0}$ is the largest; $S_{1}$ is
an enhancement over $S_{0}$ and $S_{2}$ is an enhancement over $S_{1}$.
If $p_{0},p_{1}$ and $p_{2}$ are the population fractions of numbers
starting from $1xy$ within the sets $S_{0},S_{1}$ and $S_{2}$ respectively,
we may write \begin{equation}
P_{1xy}^{1\alpha\beta}=\frac{p_{0}|S_{0}|+p_{1}|S_{1}|+p_{2}|S_{2}|}{|S_{0}|+|S_{1}|+|S_{2}|}\end{equation}
where, $|S_{j}|$ is the number of elements in $S_{j}$. Clearly,
$|S_{1}|=\frac{\alpha}{2}|S_{0}|$ and $|S_{2}|=\frac{\beta}{4}|S_{0}|$.
In $S_{0}$, the second and the third digits are equally distributed,
i.e., $100,101,110,111$ appear with equal populations. Hence $p_{0}=\frac{1}{4}$.
In $S_{1}$, all numbers have second digit $0$ and the third digit
is equally distributed between $1$ and $0$. So, $p_{1}=\delta_{x0}\frac{1}{2}$.
In $S_{2}$, all numbers have second digit $\alpha$, and third digit
$0$. Therefore, $p_{2}=\delta_{x\alpha}\delta_{y0}$. Thus, \begin{equation}
P_{1xy}^{1\alpha\beta}=\frac{1+\alpha\delta_{x0}+\beta\delta_{x\alpha}\delta_{y0}}{4+2\alpha+\beta}\end{equation}
The equation $P_{1xy}=\sum_{\alpha,\beta}P_{1xy}^{1\alpha\beta}P_{1\alpha\beta}$
reads\begin{equation}
\left[\begin{array}{cccc}
\nicefrac{1}{4} & \nicefrac{2}{5} & \nicefrac{1}{3} & \nicefrac{2}{7}\\
\nicefrac{1}{4} & \nicefrac{1}{5} & \nicefrac{1}{3} & \nicefrac{2}{7}\\
\nicefrac{1}{4} & \nicefrac{1}{5} & \nicefrac{1}{6} & \nicefrac{2}{7}\\
\nicefrac{1}{4} & \nicefrac{1}{5} & \nicefrac{1}{6} & \nicefrac{1}{7}\end{array}\right]\left[\begin{array}{c}
P_{100}\\
P_{101}\\
P_{110}\\
P_{111}\end{array}\right]=\left[\begin{array}{c}
P_{100}\\
P_{101}\\
P_{110}\\
P_{111}\end{array}\right]\end{equation}
The solution, after normalizing the sum to 1 is \[
\begin{array}{c}
P_{100}=0.3152\\
P_{101}=0.2626\\
P_{110}=0.2251\\
P_{111}=0.1969\end{array}\]
Using, $P_{100}+P_{101}=P_{10}$and $P_{110}+P_{111}=P_{11}$, we
obtain $P_{10}=0.5778$, the second approximation. As expected, it
is closer to the benford's value, $0.5849625$, than the first approximation. 

Higher order approximations can be obtained by considering a larger
number of digits. Considering $k$ digits after the first digit, the
equation to be solved is a $2^{k}\times2^{k}$ matrix equation \begin{equation}
P_{1x_{1}\cdots x_{k}}=\sum_{\{\alpha_{i}\}}P_{1x_{1}\cdots x_{k}}^{1\alpha_{1}\cdots\alpha_{k}}P_{1\alpha_{1}\cdots\alpha_{k}}\end{equation}
where $P_{1x_{1}\cdots x_{k}}$ is the probability that an unscaled
integer starts with $1x_{1}\cdots x_{k}$ and the matrix element,
$P_{1x_{1}\cdots x_{k}}^{1\alpha_{1}\cdots\alpha_{k}}$ is the corresponding
probability with a scale of $1\alpha_{1}\cdots\alpha_{k}0\cdots0$.
This can be evaluated easily. For values of $k$ up to 10, they were
solved numerically using python. Table -1 summarizes the results.
The values suggest a neat convergence to the benford's value. Interestingly,
the relative error falls exponentially. In the next section, we shall
prove it analytically. 

\begin{table}[H]
\begin{tabular}{|c|c|c|}
\hline 
$k$ & $P_{10}$ & Rel err\tabularnewline
\hline
\hline 
1 & 0.571428 & 0.023\tabularnewline
\hline 
2 & 0.577861 & 0.012\tabularnewline
\hline 
3 & 0.581339 & 0.0062\tabularnewline
\hline 
4 & 0.583135 & 0.0031\tabularnewline
\hline 
5 & 0.584045 & 0.00156\tabularnewline
\hline 
6 & 0.584503 & 0.00078\tabularnewline
\hline 
7 & 0.584732 & 0.00039\tabularnewline
\hline 
8 & 0.584847 & 0.00019\tabularnewline
\hline 
9 & 0.584905 & 0.000097\tabularnewline
\hline 
10 & 0.584933 & 0.000049\tabularnewline
\hline
\end{tabular}

\caption{Estimates of $P_{10}$ up to k=10. Value according to benford's law:
$P_{10}=0.584962$}

\end{table}

\section{Analytical Solution}

In this section, we show that the benford's law is an exact solution
to equation{[}12{]}. We are to solve the equation for $P_{1x_{1}\cdots x_{k}}$
in the limit of $k\rightarrow\infty$. And the matrix elements in
this equation are evaluated in appendix A.\begin{equation}
P_{1x_{1}\cdots x_{k}}^{1\alpha_{1}\cdots\alpha_{k}}=\frac{1+Q_{\alpha x}}{2^{k}(1+\alpha)}\end{equation}
 We are to show that the solution is logarithmic, i.e., $P_{1x_{1}\cdots x_{k}}=Log[1+\frac{1}{1x_{1}\cdots x_{k}}]$.
Observe that this function has a first approximation of $\frac{1}{1x_{1}\cdots x_{k}}$,
in the large $k$ limit. Hence, we shall first show that this is a
solution in the limit of large k. That is, we are to show, that \begin{equation}
\frac{1}{(1+x)}=\underset{k\rightarrow\infty}{lim}\sum\frac{1}{2^{k}}\frac{1+Q_{\alpha x}}{(1+\alpha)^{2}}\end{equation}
$x$ and $\alpha$ are numbers between $0$ and $1$ with $k$ places.
In the limit of $k\rightarrow\infty$, $x$ and $\alpha$ are any
real numbers between $0$ and $1$ and the sum is replaced by an integral\begin{equation}
\frac{1}{(1+x)}=\int_{0}^{1}d\alpha\frac{1+Q_{\alpha x}}{(1+\alpha)^{2}}\end{equation}
We are to show the above relation. $Q_{\alpha x}$ is the sum of an
infinite sereis. The integral is easily evaulated for each of these
terms, and then summed up. The details of this proof has been completed
in appendix B. 

For a finite value of k, to evaluate $P_{1\beta_{1}\cdots\beta_{k}}$,
we write it as \begin{equation}
P_{1\beta_{1}\cdots\beta_{k}}=\sum_{\{\alpha_{i}\}}P_{1\beta_{1}\cdots\beta_{k}\alpha_{1}\cdots\alpha_{l}}\end{equation}
We have shown that in the large $l$ limit, \begin{equation}
\underset{l\rightarrow\infty}{lim}P_{1\beta_{1}\cdots\beta_{k}\alpha_{1}\cdots\alpha_{l}}=\frac{1}{1\beta_{1}\cdots\beta_{k}\alpha_{1}\cdots\alpha_{l}}\end{equation}
Thus,\begin{eqnarray*}
P_{1\beta_{1}\cdots\beta_{k}}= & \underset{l\rightarrow\infty}{lim}\sum_{\{\alpha_{i}\}}\frac{1}{1\beta_{1}\cdots\beta_{k}\alpha_{1}\cdots\alpha_{l}}= & \underset{l\rightarrow\infty}{lim}\sum_{n=0}^{2^{l}}\frac{1}{2^{l}(1\beta_{1}\cdots\beta_{k})+n}\\
 &  & =Ln\left(\frac{1\beta_{1}\cdots\beta_{k}+1}{1\beta_{1}\cdots\beta_{k}}\right)\end{eqnarray*}
Normalizing, we obtain the benford's law \begin{equation}
P_{1\beta_{1}\cdots\beta_{k}}=Log_{2}\left(\frac{1\beta_{1}\cdots\beta_{k}+1}{1\beta_{1}\cdots\beta_{k}}\right)\end{equation}

\section{Discussion}

So far, little light has been thrown in to the counterintuitive nature
of benford's law. We haven't reconstructed our intuition so as to
understand the law. The origin of the anomalous behaviour is still
unclear. A strong reason why it is counterintuitive is that, the cardinalities
of numbers starting from any digit is the same, and therefore we expect
the probabilities to be the same as well. One step towards understanding
it is to realise that, the probabilities measure the $occurances$
and not the $cardinalities$. 

To understand it better, let $\{a_{i}\}$ be a sequence and $\{b_{i}\}$
be a sub sequence of $\{a_{i}\}$. For instance, let $a_{i}=i$ and
$b_{i}=2i$. $a_{i}$ is the sequence of positive integers and $b_{i}$
is the subsequence of even numbers. The probability that a randomly
chosen element in $\{a_{i}\}$ is also an element in $\{b_{i}\}$
is $\frac{1}{2}$. Now, let $\{c_{i}\}$ be a subsequence of $\{b_{i}\}$,
$c_{i}=4i$, the sequence of multiples of four. The probablity that
a randomly chosen element in $\{a_{i}\}$ is also an element in $\{c_{i}\}$
is $\frac{1}{4}$. Even though $\{b_{i}\}$ and $\{c_{i}\}$ have
the same cardinalities, and can be mapped to each other, the probabilities
are not equal. In fact, the sequence $\{a_{i}\}$ can be rearranged
such that every alternate term is an element of $\{c_{i}\}$. \[
\{a'_{i}\}:1,4,2,8,3,12,5,16,\cdots\]
This sequence $\{a'_{i}\}$ is a rearrangement of $\{a_{i}\}$. The
probability that a random element belongs to $\{c_{i}\}$ is now $\frac{1}{2}$.
Hence, this probability is unrelated to the cardinality; instead,
it is a measure of $frequency$ $of$ $occurance$ of the elements
of $\{c_{i}\}$ in the parent sequence $\{a_{i}\}$. Hence, it changes
on rearranging the parent sequence. 

In the above examples, all the occurances were periodic. Thus, even
though the sequences were infinite, due to the periodicity, the calculation
of the probability was as simple as it is in case of a finite set.
However, in a benford sequence, there is no such periodicity, and
therefore, the calculation is nontrivial. In this paper, we have outlined
a possible analytical explanation for benford's law for base 2. It
is very likely that, a similar strategy can yeild the law for any
base. Therefore, further work in this direction is expected to be
fruitful.

\section{Appendices}

\subsection{Appendix A: Evaluating the matrix elements}

In this appendix, we evaluate the coefficients $P_{1x_{1}\cdots x_{k}}^{1\alpha_{1}\cdots\alpha_{k}}$.
It is the population fraction of numbers starting from $1x_{1}\cdots x_{k}$
in the set $S=[0,1\alpha_{1}\cdots\alpha_{k}000\cdots0)$. We shall
use the same strategy again: break this set in to disjoint chunks.
\[
S=[0,100\cdots0)\cup[100\cdots0,1\alpha_{1}0\cdots0)\cup\cdots\cup[1\alpha_{1}\cdots\alpha_{k-1}00\cdots0,1\alpha_{1}\cdots\alpha_{k}0\cdots0)\]
defining the sets,\begin{eqnarray*}
S_{0}=[0,100\cdots0) & \& & S_{r}=[1\alpha_{1}\cdots\alpha_{r-1}00\cdots0,1\alpha_{1}\cdots\alpha_{r-1}\alpha_{r}0\cdots0)\end{eqnarray*}
 we may write \[
S=S_{0}\cup S_{1}\cup\cdots\cup S_{k}\]
Writing $p_{r}=$population fraction of numbers starting from $1x_{1}\cdots x_{k}$
in the set $S_{r}$ and $|S_{r}|$=size of $S_{r}$, we may write
\[
P_{1x_{1}\cdots x_{k}}^{1\alpha_{1}\cdots\alpha_{k}}=\frac{p_{0}|S_{0}|+p_{1}|S_{1}|+\cdots+p_{k}|S_{k}|}{|S_{0}|+|S_{1}|+\cdots+|S_{k}|}\]
since the sets are disjoint. Clearly, $|S_{r}|=\frac{\alpha_{r}}{2^{r}}|S_{0}|$.
And, in $|S_{0}|$, all numbers are equally populated, thus, $p_{0}=\frac{1}{2^{k}}$.
In the set $S_{r}$, all numbers have the first $r-1$ digits equal
to $\alpha_{1}\cdots\alpha_{r-1}$ respectively, and the $r^{th}$
digit is zero. The rest of the$k-r$ digits are 0 or 1 with a probabililty
of $\frac{1}{2}$ each. Thus, \[
p_{r}=\delta_{\alpha_{1}x_{1}}\delta_{\alpha_{2}x_{2}}\cdots\delta_{\alpha_{r-1}x_{r-1}}\delta_{0x_{r}}.\frac{1}{2^{k-r}}\]
Therefore, \[
P_{1x_{1}\cdots x_{k}}^{1\alpha_{1}\cdots\alpha_{k}}=\frac{1+\alpha_{1}\delta_{0x_{1}}+\alpha_{2}\delta_{0x_{2}}\delta_{\alpha_{1}x_{1}}+\cdots+\alpha_{k}\delta_{0x_{k}}\delta_{\alpha_{k-1}x_{k-1}}\cdots\delta_{\alpha_{1}x_{1}}}{1\alpha_{1}\cdots\alpha_{k}}\]
where $1\alpha_{1}\cdots\alpha_{k}=2^{k}+\alpha_{1}2^{k-1}+\cdots+\alpha_{k}$.
We can express it conveneintly in a better notation. Let us define\begin{eqnarray*}
\alpha=\frac{\alpha_{1}}{2}+\frac{\alpha_{2}}{2^{2}}+\cdots+\frac{\alpha_{k}}{2^{k}} & \& & x=\frac{x_{1}}{2}+\frac{x_{2}}{2^{2}}+\cdots+\frac{x_{k}}{2^{k}}\end{eqnarray*}
 $\alpha$ and $x$ are numbers between $0$ and $1$ with k places.
In this notation, \[
P_{1x_{1}\cdots x_{k}}^{1\alpha_{1}\cdots\alpha_{k}}=\frac{1+\alpha_{1}\delta_{0x_{1}}+\alpha_{2}\delta_{0x_{2}}\delta_{\alpha_{1}x_{1}}+\cdots+\alpha_{k}\delta_{0x_{k}}\delta_{\alpha_{k-1}x_{k-1}}\cdots\delta_{\alpha_{1}x_{1}}}{2^{k}(1+\alpha)}\]
also, for brevity, define \[
\alpha_{1}\delta_{0x_{1}}+\alpha_{2}\delta_{0x_{2}}\delta_{\alpha_{1}x_{1}}+\cdots+\alpha_{k}\delta_{0x_{k}}\delta_{\alpha_{k-1}x_{k-1}}\cdots\delta_{\alpha_{1}x_{1}}=Q_{\alpha x}\]
so that \[
P_{1x_{1}\cdots x_{k}}^{1\alpha_{1}\cdots\alpha_{k}}=\frac{1+Q_{\alpha x}}{2^{k}(1+\alpha)}\]

\subsection{Appendix B: Analytical Solution}

In this appendix, we show that\[
\frac{1}{(1+x)}=\int_{0}^{1}d\alpha\frac{1+Q_{\alpha x}}{(1+\alpha)^{2}}\]
Note that the first term, after performing the integral is $\frac{1}{2}$.
For conveneince, let us make the substitution $t=1-x;$ $t_{r}=1-x_{r}$.
The integral corresponding to $r^{th}$ term in $Q_{\alpha x}$ is
given by \[
\int_{0}^{1}\frac{d\alpha}{(1+\alpha)^{2}}t_{r}\alpha_{r}\delta_{\alpha_{1}x_{1}}\delta_{\alpha_{2}x_{2}}\cdots\delta_{\alpha_{r-1}x_{r-1}}\]
$t_{r}$ can be taken out. The delta terms inside fix the first $r$
places of $\alpha$. $\alpha_{i}=x_{i}=1-t_{i}$ up to $i=r-1$ and
$\alpha_{r}=1$. Thus, the integral can be written as:\[
t_{r}\int_{a_{r}}^{b_{r}}\frac{d\alpha}{(1+\alpha)^{2}}=t_{r}\left(\frac{1}{(1+a_{r})}-\frac{1}{(1+b_{r})}\right)\]
where, $[a_{r},b_{r}]$ is the range in which none of the deltas inside
are zero. This range is given by \[
a_{r}=0.x_{1}x_{2}\cdots x_{r-1}1=x^{[r-1]}+\frac{1}{2^{r}}=1-t^{[r-1]}-\frac{1}{2^{r}}\]
and \[
b_{r}=0.x_{1}x_{2}\cdots x_{r-1}1111\cdots=1-t^{[r-1]}\]
where $t^{[r]}$ is the approximation of $t$ up to $r$ places; \[
t^{[r]}=\frac{t_{1}}{2}+\frac{t_{2}}{2^{2}}+\frac{t_{3}}{2^{3}}+\cdots+\frac{t_{r}}{2^{r}}\]
Thus, the integral corresponding to the $r^{th}$ term in $Q_{\alpha x}$
is \[
t_{r}\left(\frac{\frac{1}{2^{r}}}{(2-t^{[r-1]})(2-t^{[r-1]}-\frac{1}{2^{r}})}\right)\]
Thus, summing up, we obtain \[
\sum\frac{1}{2^{k}}\frac{1+Q_{\alpha x}}{(1+\alpha)^{2}}=\frac{1}{2}+\sum_{r=1}^{\infty}\frac{t_{r}}{2^{r}}\left(\frac{1}{(2-t^{[r-1]})(2-t^{[r-1]}-\frac{1}{2^{r}})}\right)\]
Next we show that the sereis on the RHS sums up to $\frac{1}{1+x}$
or $\frac{1}{2-t}$. Consider, \[
\frac{1}{2-t^{[r+1]}}-\frac{1}{2-t^{[r]}}=\frac{t^{[r+1]}-t^{[r]}}{(2-t^{[r]})(2-t^{[r+1]})}=\frac{t_{r+1}}{2^{r+1}}\frac{1}{(2-t^{[r]})(2-t^{[r]}-\frac{t_{r+1}}{2^{r+1}})}\]
Using the above repeatedly, we may expand $\frac{1}{2-t^{[r]}}$ as
\[
\frac{1}{2-t^{[r]}}=\frac{1}{2}+\frac{t_{1}}{2}\frac{1}{2(2-\frac{t_{1}}{2})}+\frac{t_{2}}{2^{2}}\frac{1}{(2-\frac{t_{1}}{2}-\frac{t_{2}}{4})(2-\frac{t_{1}}{2}-\frac{t_{2}}{4}-\frac{t_{3}}{8})}+\cdots+\frac{t_{r}}{2^{r}}\frac{1}{(2-t^{[r-1]})(2-t^{[r-1]}-\frac{t_{r}}{2^{r}})}\]
Further, since, $t_{r}$ can take only two values, $0$ and $1$,
we may write \[
\frac{t_{r}}{2^{[r]}}\frac{1}{(2-t^{[r-1]})(2-t^{[r-1]}-\frac{t_{r}}{2^{r}})}=\frac{t_{r}}{2^{r}}\frac{1}{(2-t^{[r-1]})(2-t^{[r-1]}-\frac{1}{2^{r}})}\]
Thus, \[
\frac{1}{2-t^{[r]}}=\frac{1}{2}+\frac{t_{1}}{2}\frac{1}{2(2-\frac{1}{2})}+\frac{t_{2}}{2^{2}}\frac{1}{(2-\frac{t_{1}}{2}-\frac{1}{4})(2-\frac{t_{1}}{2}-\frac{t_{2}}{4}-\frac{1}{8})}+\cdots+\frac{t_{r}}{2^{r}}\frac{1}{(2-t^{[r-1]})(2-t^{[r-1]}-\frac{1}{2^{r}})}\]
And continuing the sereis, \[
\frac{1}{2-t}=\frac{1}{2}+\sum_{r=1}^{\infty}\frac{t_{r}}{2^{r}}\left(\frac{1}{(2-t^{[r-1]})(2-t^{[r-1]}-\frac{1}{2^{r}})}\right)\]
Thus, \[
\int_{0}^{1}d\alpha\frac{1+Q_{\alpha x}}{(1+\alpha)^{2}}=\frac{1}{2-t}=\frac{1}{(1+x)}\]

\end{document}